\documentclass[a4paper,11pt]{article}
\usepackage{subcaption}
\usepackage{epsfig, graphicx, subfloat, float}
\usepackage{latexsym}
\usepackage{cite}
\usepackage{amssymb}
\usepackage{latexsym}
\usepackage{mathrsfs}
\usepackage{amsmath}
\title{Constant-roll, cosmic acceleration, and massive neutrinos}
\author{V. Anari\footnote{v.anari@ut.ac.ir} \,and H. Mohseni Sadjadi\footnote{mohsenisad@ut.ac.ir}
\\ {\small Department of Physics, University of Tehran,}
\\ {\small }
\\ {\small P. O. B. 14395-547, Tehran 14399-55961, Iran}}

\begin{document}
\maketitle
\begin{abstract}
We propose a model to describe the late-time cosmic acceleration in the context of the constant-roll model. By considering a coupling between massive neutrinos and the quintessence, the onset of evolution of the quintessence is related to the neutrinos' behavior. When the neutrinos become non-relativistic, the dark energy density increases from zero and results in late-time cosmic acceleration alleviating the coincidence problem. The quintessence potential is such that it evolves under the constant-roll condition giving rise to persistent late-time acceleration.
\end{abstract}

\section{Introduction}
For more than two decades, astrophysical data have shown that the Universe's expansion is accelerating \cite{acc1,acc2,acc3}.
Different models have been used to investigate the origin of this acceleration. A straightforward method to describe this phenomenon is introducing a cosmological constant, albeit this model suffers from fine-tuning and coincidence problems \cite{wein1,wein2,wein3}. In addition, it seems that there is a significant discrepancy between the Hubble parameter values derived by fitting this model to CMB (cosmic microwave background) data and the value obtained locally by using distance indicators \cite{ten1,ten2,ten3,ten4,ten5}.

Another category of dark energy models is the dynamical dark energy, whose a simple candidate is a scalar field \cite{quint1,quint2,quint3,quint4}.
In the context of the scalar-tensor models, this field may be hidden from the local tests in dense regions, where the matter density is more than a critical value. In this framework, the conformal coupling of the scalar field to the matter may trigger the acceleration of the Universe at a specific redshift, when the matter density is diluted below the critical density, by activating the dark energy component. This could also happen when the massive neutrinos became non-relativistic during the Universe evolution, alleviating the coincidence problem. Conformal coupling of neutrinos to a scalar field was also used to introduce an early dark energy model to describe the Hubble tension problem \cite{screen1,screen2}.

The scalar field models were also employed to study the inflation in the early universe. To provide enough e-folds one can consider a scalar field rolling down slowly its potential. Appropriate choice of the potential makes the slow-roll inflationary model consistent with observation \cite{planckin1,planckin2}. It is also possible to consider more general models like the constant-roll \cite{cr-2,constant-roll,cr-1,cr0}, which is obtained by replacing the slow-roll condition by $\ddot{\phi}(t)=\beta H\dot{\phi}(t)$ for the scalar field ($\phi$), where $\beta$ is a constant and $H$ is the Hubble parameter. In this framework, one may describe non-Gaussianity in the CMB spectrum. This assumption determines the form of the scalar field potential. Analytic solutions are available in this situation \cite{cr1,cr2,cr3,cr4}. By considering a slow-roll quintessence, generalization of the slow-roll to the late time was proposed in \cite{shib1}, resulting in thawing and freezing models \cite{shib2,shib3}.

In this article, inspired by the constant-roll model and conformally coupled neutrinos to the quintessence, we propose a model that can describe the onset of positive late-time cosmic acceleration. In our model, the dark energy density is zero initially. Due to the change of the neutrino behavior from relativistic to non-relativistic, the $Z_2$ symmetry is broken and the evolution of the quintessence begins. As the Universe expands and the total density decreases, the quintessence evolves under the constant-roll condition, which leads to persistent late-time acceleration.

The scheme of this paper is as follows: in the next section we construct our model and explain how the dark energy can increase from zero energy density and its evolution under the constant-roll condition can lead to a persistent late-time acceleration. We then illustrate and discuss our results with numerical examples for two sets of parameters. Finally, we conclude our results in the last section.

Throughout this paper we use units $\hbar=c=8\pi G=1$ and metric signature (-,+,+,+).

\section{Dark Energy-Neutrinos Coupling and Cosmic Acceleration with a Constant Rate of Roll}
We adopt the action\cite{action}
\begin{equation}\label{action}
S=\int{\left[ \frac{R}{2}-\frac{1}{2}\partial _{\mu}\phi \partial ^{\mu}\phi -V\left( \phi \right) \right] \sqrt{-g}d^4x}+S_\nu\left[ \tilde{g}_{\mu \nu} \right] +S_m+S_r,
\end{equation}
where $S_m$ and $S_r$ are actions for the pressure-less matter and relativistic matter, respectively. Also, $S_\nu$ is the action for neutrinos which interact with the scalar field $\phi$. We assume that this interaction is via a conformal coupling in the neutrino sector
\begin{equation}\label{A(phi)}
\tilde{g}_{\mu \nu}=A^2\left( \phi \right) g_{\mu \nu} \, ,
\end{equation}
where $A(\phi)$ is a positive function. We must note that these neutrinos are not necessarily ordinary neutrinos that have electroweak interactions, and all or a portion of ordinary neutrinos can contribute to either the relativistic matter action $S_r$ or the pressure-less matter action $S_m$. The quintessence-neutrino interactions could also be introduced in the context of mass varying neutrinos without using the conformal coupling method \cite{fardon, MaVaNs}.\\

The Universe is taken as a spatially flat Friedmann-Lemaître-Robertson-Walker (FLRW) space-time
\begin{equation}
ds^2=-dt^2+a^2(t)\left(dx^2+dy^2+dz^2\right),
\end{equation}
where $a(t)$ is the scale factor. Variation of (\ref{action}) with respect to the metric leads to Friedman equations
\begin{equation}\label{Friedman1}
H^2=\left(\dfrac{\dot{a}}{a}\right)^2=\frac{1}{3}\sum_i\rho_i ,
\end{equation}
\begin{equation}\label{Friedman2}
\dot{H}+H^2=\dfrac{\ddot{a}}{a}=-\frac{1}{6}\sum_i{(\rho_i+ 3P_i)} ,
\end{equation}
where an overdot denotes derivative with respect to the cosmic time $t$, $H$ is the Hubble parameter and $\rho_i$ and $P_i$ are the energy density and pressure of the $i$th cosmic fluid, respectively. Note that $P_m=0$, $P_r=\dfrac{1}{3}\rho_r$ and the energy density and pressure of the scalar field are respectively given by
\begin{equation}\label{rho_phi}
\rho_\phi = \frac{1}{2}\dot{\phi}^2+V(\phi) ,
\end{equation}
\begin{equation}\label{P_phi}
P_\phi =\frac{1}{2}\dot{\phi}^2-V(\phi) .
\end{equation}
For the Universe to be positively accelerated, the condition $\ddot{a}>0$ must be satisfied; which yields
\begin{equation}
V(\phi)>\dot{\phi}^2+\frac{1}{2}\left(\rho _\nu +3P_\nu +\rho _m+2\rho _r\right) .
\end{equation}
Moreover, variation of (\ref{action}) with respect to $\phi$, gives the equation of motion of the scalar field
\begin{equation}\label{EOM1}
\ddot{\phi}+3H\dot{\phi}+V_{\text{eff.},\phi}=0 .
\end{equation}
where
\begin{equation}
V_{\text{eff.},\phi}(\phi)=V_{,\phi}(\phi)+ \frac{A_{,\phi}(\phi)}{A(\phi)}(\rho_\nu - 3P_\nu)
\end{equation}
and $\{ _{,\phi}\}$ denotes derivative with respect to the scalar field $\phi$.\\

The continuity equations are given by
\begin{equation}\label{CEnu}
\dot{\rho}_\nu +3H(\rho_\nu +P_\nu)=\frac{A_{,\phi}(\phi)}{A(\phi)}(\rho_\nu - 3P_\nu)\dot{\phi},
\end{equation}
\begin{equation}
\dot{\rho}_m+3H\rho_m=0,
\end{equation}
\begin{equation}
\dot{\rho}_r+4H\rho _r=0 .
\end{equation}
If the neutrino masses are $\phi$ dependent and they are in thermal equilibrium, by using Fermi-Dirac distribution, one finds \cite{Brookfield, Peccei}
\begin{equation}\label{CEmvnu}
\dot{\rho}_\nu +3H(\rho_\nu +P_\nu)=\frac{m_{\nu,\phi}(\phi)}{m_\nu(\phi)}(\rho_\nu - 3P_\nu)\dot{\phi}.
\end{equation}

Therefore, taking $A(\phi)=\dfrac{m_\nu(\phi)}{m_\nu^*}$ in (\ref{CEnu}),  results in (\ref{CEmvnu}), where $m_\nu^*$ is a mass scale. In other words, the two approaches, i.e. mass varying neutrinos whose masses depend on the scalar field $\phi$ \cite{Pettorino, Brookfield, Peccei}, and conformal coupling in neutrino sector through a function of $\phi$ \cite{MaVaNs}, give the same equations of motion.\\

Now to construct our model, we require that
\begin{enumerate}
\item[(I)] Initially, when neutrinos are relativistic ($m_\nu \ll T_\nu$), $\rho_\phi$ is negligible and the Universe is in a decelerated phase.
\item[(II)] As the Universe expands and temperature decreases, neutrinos exit from the relativistic state, the $Z_2$ symmetry breaks and the evolution of the scalar field begins.
\item[(III)] As the energy densities $\rho_\nu$, $\rho_m$ and $\rho_r$ decrease, the scalar field evolves under constant-roll conditions, i.e. $\ddot{\phi}=\beta H\dot{\phi}$, which leads to persistent late-time cosmic acceleration.
\end{enumerate}

$A(\phi)$ and $V(\phi)$ must be chosen such that the above conditions are satisfied. Thus, we assume that $V(\phi)$ and $A(\phi)$ have $Z_2$ symmetry. Initially, when neutrinos are relativistic, we assume that the scalar field is at the stable point $\phi=0$, where  $V(\phi)=0$. In this era, since $V(\phi)=0$ and $\dot{\phi}=0$, according to (\ref{rho_phi}), the dark energy density vanishes. Also, since $\rho_\nu\approx3P_\nu$, we have $V_{\text{eff.}}(\phi)=V(\phi)$ and the interaction in (\ref{CEnu}) is non-operative. In order to have an initial stable solution, it is required that the potential satisfies the below conditions
\begin{eqnarray}\label{condition}
V_{,\phi}(0)=0 ,\nonumber\\
V_{,\phi\phi}(0)>0 .
\end{eqnarray}

With the expansion of the Universe, the neutrinos exit from the relativistic phase such that ($\rho_\nu-3P_\nu$) becomes significant. Whenever\footnote{According to the $Z_2$ symmetry of $A(\phi)$, $A_{,\phi}(0)=0$.}
\begin{equation}
V_{,\phi\phi}(0)+\frac{A_{,\phi\phi}(0)}{A(0)}(\rho_\nu-3P_\nu)<0 ,
\end{equation}
the effective potential at $\phi=0$ becomes concave and this point becomes unstable. Thus, the scalar field rolls down the effective potential and the $Z_2$ symmetry is broken. In contrast to the effective potential, the potential is convex at $\phi=0$ and the scalar field climbs its own potential, giving rise to a non-zero dark energy density. This mechanism can provide the positive potential required for cosmic acceleration. Using this mechanism and applying the slow-roll conditions, a non-persistent late-time cosmic acceleration is introduced in \cite{MaVaNs}.\\

In this paper, we introduce a persistent late-time cosmic acceleration with a constant rate of roll.
When the energy densities $\rho_\nu$, $\rho_m$ and $\rho_r$ have decreased significantly, they can be neglected and the eqs. (\ref{Friedman1}), (\ref{Friedman2}) and (\ref{EOM1}) become similar to the inflation era. Hence, in order for the condition (III) to be satisfied, $V(\phi)$ must be chosen similar to the potentials which are used in the constant-roll inflationary models. The most general potential that can result in $\ddot{\phi}=\beta H\dot{\phi}$ in the inflation era is \cite{constant-roll}
\begin{equation}
W(\phi)=(\beta +3)C^2_1e^{\sqrt{-2\beta}\phi}+2(3-\beta)C_1C2+(\beta +3)C^2_2e^{-\sqrt{-2\beta}\phi},
\end{equation}
where $C_1$ and $C_2$ are constant. Therefore, we can choose the potential of our model to be
\begin{equation}\label{potential}
V(\phi)=f(\phi)W(\phi),
\end{equation}
where $f(\phi)$ is an even function such that $f(0)=0$ and as $\phi$ is displaced from zero, $f(\phi)\approx1$. A possible choice for $f(\phi)$ is
\begin{equation}
f(\phi)=1-e^{-\gamma \phi^2},
\end{equation}
where $\gamma$ is a positive constant. The only choice for $W(\phi)$ that has $Z_2$ symmetry and can lead to persistent late-time acceleration is\footnote{If $\phi>0$, we choose $C_1=0$ and $C_2=M$. Otherwise, we choose $C_1=M$ and $C_2=0$.}
\begin{equation}
W(\phi)=(\beta +3)M^2e^{-\sqrt{-2\beta}\mid\phi\mid},
\end{equation}
where $-1<\beta <0$.\footnote{The lower limit is resulted from the requirement that the late-time acceleration be persistent.}\\

\begin{figure}[H]
	\centering
	\includegraphics[height=6 cm]{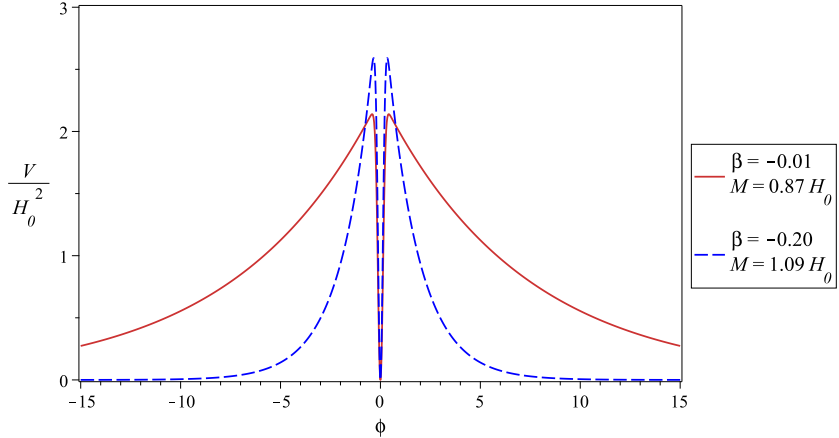}
	\caption{The potential (\ref{potential}) with respect to $\phi$ for $\gamma=30$.}
	\label{fig_V}
\end{figure}

Now to show how the model works, we illustrate our results with a numerical example for two sets of parameters $\{\beta, M\}$. The potential (\ref{potential}) is depicted for these two sets and $\gamma=30$ in figure \ref{fig_V}.

\begin{figure}[H]
	\centering
	\includegraphics[height=6 cm]{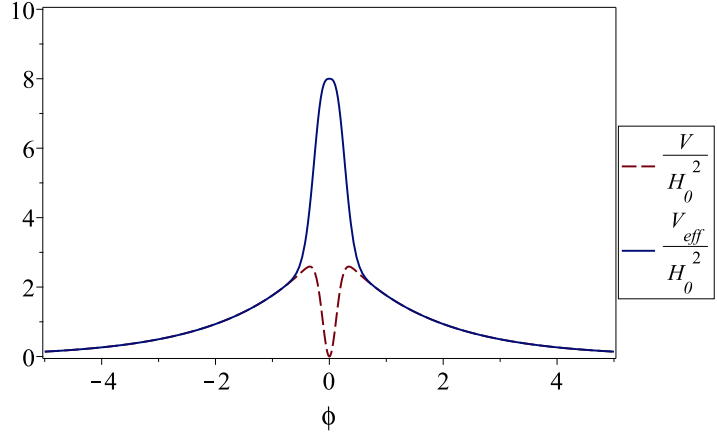}
	\caption{The effective potential (\ref{V_eff}) with respect to $\phi$ for $\gamma=30$.}
	\label{fig_V_eff}
\end{figure}

Similar to \cite{MaVaNs,Mota,Pietroni,JCAP,JCAP1}, we choose $A(\phi)=e^{-\alpha\phi^2}$ which satisfies all the conditions mentioned earlier in this paper, and where $\alpha$ is a positive constant. In order to investigate the evolution of the Universe, we must solve the below equations simultaneously.
\begin{eqnarray}\label{set of eqs}
&&\dot{a}=Ha, \nonumber\\
&&\dot{\rho}_\nu =-3H(\rho_\nu +P_\nu)+\frac{A_{,\phi}(\phi)}{A(\phi)}(\rho_\nu - 3P_\nu)\dot{\phi}, \nonumber\\
&&\dot{\rho}_m=-3H\rho_m ,  \nonumber\\
&&\dot{\rho}_r=-4H\rho _r , \nonumber\\
&&\ddot{\phi}=-3H\dot{\phi}-V_{,\phi}(\phi)- \frac{A_{,\phi}(\phi)}{A(\phi)}(\rho_\nu - 3P_\nu) ,\nonumber\\
&&H^2=\frac{1}{3}( \frac{1}{2}\dot{\phi}^2+V(\phi) +\rho_\nu+\rho_m+\rho_r).
\end{eqnarray}
In the presence of $P_\nu$, (\ref{set of eqs}) becomes very complicated to solve, even numerically. Hence, similar to \cite{MaVaNs} we set the beginning time of our numerical study to be when the neutrinos are completely non-relativistic, i.e. $P_\nu\ll\rho_\nu$. Thus, we ignore $P_\nu$ in (\ref{set of eqs}). Accordingly, the effective potential for the scalar field is
\begin{equation}\label{V_eff}
V_{eff.}(\phi)=V(\phi)+m_\nu(\phi) n_\nu .
\end{equation}
In figure \ref{fig_V_eff}, the potential (\ref{potential}) and the effective potential (\ref{V_eff}) are depicted for $\alpha=13$, $\beta=-0.20$, $\gamma=30$, $M=1.09\,H_0^2$ and $m_\nu^* n_\nu=8\,H_0^2$.

We choose two sets of parameters $\{\beta=-0.01, M=0.87 H_0\}$ and $\{\beta=-0.20, M=1.09 H_0\}$, the other parameters of the model to be
\begin{equation}\label{parameters}
\alpha=13, \qquad \gamma=30 .
\end{equation}
and the following initial conditions\footnote{One must note that according to (\ref{set of eqs}), $\rho_m$ and $\rho_r$ can be obtained with respect to $a$ and are $\rho_m=\rho_m^0a^{-3}$ and $\rho_r=\rho_r^0a^{-4}$ where $\rho_m^0$ and $\rho_r^0$ are the corresponding energy densities at present time ($a=1$).}
\begin{eqnarray}\label{ICs}
&& a=\frac{1}{2000+1}, \qquad \phi=10^{-8}, \qquad \dot{\phi}=10^{-5} H_0, \nonumber\\
&& \rho_r = 4.33 \times 10^9 H_0^2, \quad \rho_m = 7.40 \times 10^9 H_0^2, \quad \rho_\nu = 1.48\times 10^9 H_0^2, \,\,\qquad\,
\end{eqnarray}
where $H_0$ is the present Hubble parameter, i.e. the Hubble parameter at $a=1$. We define dimensionless time $\tau$ as $\tau=tH_0$ and the initial conditions are set at $\tau=0$ which corresponds to the redshift is $z = 2000$, i.e. after matter-radiation equality. It is worth noting that the scalar field began its motion just after symmetry breaking where $P_\nu$ was not completely negligible (at $\tau<0$). Also, the usual expression describing neutrino-photon density ratio after $e^+ e^-$ annihilation, is given by:
\begin{equation}\label{rho_nu2}
\frac{\rho_{\nu *}}{\rho_\gamma}=\frac{7}{8}\left(\frac{4}{11}\right)^\frac{4}{3}N_{eff.},
\end{equation}
where $\rho_{\nu *}$ is the energy density of ordinary neutrinos, i.e. those neutrinos that take part in electroweak interactions, $\rho_\gamma$ is the energy density of photons,  and $N_{eff.}$ is the \textit{effective} number of neutrino species in the Universe.
Now if one assumes that only a fraction of ordinary neutrinos, $x$, which becomes non-relativistic interact with the scalar field $\phi$, and the rest of them contribute to the radiation sector. We have the following equations:
\begin{eqnarray}\label{sys_eqs}
x\rho_{\nu *}=\rho_\nu , \nonumber\\
(1-x)\rho_{\nu *}+\rho_\gamma=\rho_r .
\end{eqnarray}
Solving (\ref{rho_nu2}) and (\ref{sys_eqs}) for $x$ and $\rho_\gamma$, we have:
\begin{eqnarray}\label{sys_sol}
&&x= \frac{\rho_\nu}{\rho_\nu+\rho_r}\left(1+\frac{1}{0.227N_{eff.}}\right) , \nonumber\\
&&\rho_\gamma = \frac{\rho_\nu+\rho_r}{1+0.227N_{eff.}} .
\end{eqnarray}
According to (\ref{ICs}) and if $N_{eff}\simeq3.046$:
\begin{eqnarray}
&&x= 62\% , \nonumber\\
&&\rho_\gamma = 3.43 \times 10^9 H_0^2 .
\end{eqnarray}
\\
Moreover, if conformaly coupled neutrinos exited the relativistic phase at $z=z_{nr}$, their temperature $T_\nu^*$ at that time, in terms of the present CMB temperature $T_\gamma^0$, was
\begin{equation}\label{T_nu}
T_\nu^*=\left(\frac{4}{11}\right)^\frac{1}{3}T_\gamma^0(1+z_{nr}).
\end{equation}
At this time  the neutrino mass is  $m^*_\nu=3.15 T_\nu^*$ \cite{Pastor}, therefore,
\begin{equation}\label{m_nu}
m^*_\nu=3.15 \left(\frac{4}{11}\right)^\frac{1}{3}T_{\gamma}^0(1+z_{nr})= (1+z_{nr})=0.527\times10^{-5}(1+z_{nr})\text{ eV}\, .
\end{equation}
{\it{Note that this is true only when the considered neutrinos be in thermal equilibrium with photons in an earlier stage.}}

After becoming nonrelativisitc, due to the interaction with the scalar field, the neutrino mass varies as
\begin{equation}
m_\nu (\phi)=m^*_\nu e^{-\alpha \phi^2}.
\end{equation}
Assuming that the activation of the scalar field occurs long after the big bang nucleosynthesis (BBN), (to give rise to the late time acceleration without affecting BBN, structure formation, and so on): $z_{nr}\ll 10^9$, we have  $m^*_\nu\ll 0.527\times 10^4 \text{ eV}$.
In our numerical example, this occurs around $z\simeq2000$, we have $m^*_\nu\simeq 0.0105 \text{ eV}$. This is in agreement with the sum of neutrino masses $\Sigma m_\nu<0.12 \text{ eV}$ reported in \cite{Planck 2018}.

Note that one can also consider conformally coupled sterile neutrino. In this situation, the equation (\ref{T_nu}) does not hold unless we assume that the sterile neutrino is mixed with the active ones, and had reached thermal equilibrium with them in the early universe \cite{Gariazzo}. In \cite{Feng}, based on the Planck 2018 release and by using the CMB+BAO+SN data combination, an upper bound for the effective mass of sterile neutrinos in the $wCDM$ model is derived as $m_\nu<0.651 \text{ eV}$.
\\

The relative densities defined by $\Omega_i=\dfrac{\rho_i}{3H^2}$ are derived from (\ref{ICs}) for both $\{\beta, M\}$ sets as
\begin{equation}\label{ICs_omega}
\Omega_r = 0.328, \quad \Omega_m = 0.560, \quad \Omega_\nu = 0.112, \quad \Omega_\phi = 3.78\times 10^{-21},
\end{equation}
and the Hubble parameter is $H=6.63\times 10^{4}H_0$. The initial values chosen for $\phi$ and $\dot{\phi}$ result in  $\Omega_\phi= 3.78\times 10^{-21}$, which gives a negligible contribution in the total density.

\begin{figure}[H]
	\centering
	\includegraphics[height=6cm]{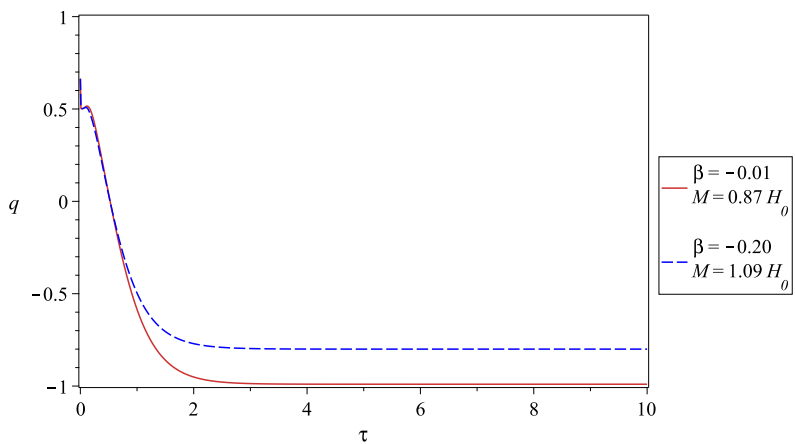}
	\caption{The deceleration parameter in terms of $\tau$, for the parameters (\ref{parameters}) and initial conditions (\ref{ICs}).}
	\label{fig_q}
\end{figure}

The deceleration parameter $q=-\dfrac{a\ddot{a}}{\dot{a}^2}$, in terms of the Hubble parameter is given by
\begin{equation}
q=-\left(1+\frac{\dot{H}}{H^2}\right).
\end{equation}
$q<0$ results in a positively accelerated universe.\\
In figure \ref{fig_q}, $q$ is plotted in terms of $\tau$. As one can see, the Universe is transited from a deceleration epoch to an acceleration epoch at a time less than the Hubble time. This acceleration for $\{\beta=-0.01, M=0.87 H_0\}$ begins at the redshift $z\simeq0.60$ and for $\{\beta=-0.20, M=1.09 H_0\}$ begins at the redshift $z\simeq0.59$ and is persistent in both cases. As the Universe expands, $q$ tends to $-1-\beta$. Thus, a universe with smaller $\lvert\beta\lvert$ is more accelerated.

\begin{figure}[H]
	\centering
	\includegraphics[height=6cm]{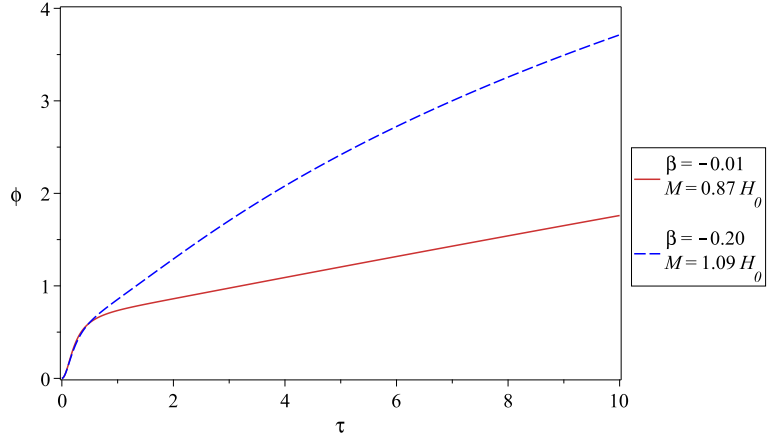}
	\caption{The scalar field in terms of $\tau$, for the parameters (\ref{parameters}) and initial conditions (\ref{ICs}).}
	\label{fig_phi}
\end{figure}

\begin{figure}[H]
	\centering
	\includegraphics[height=6cm]{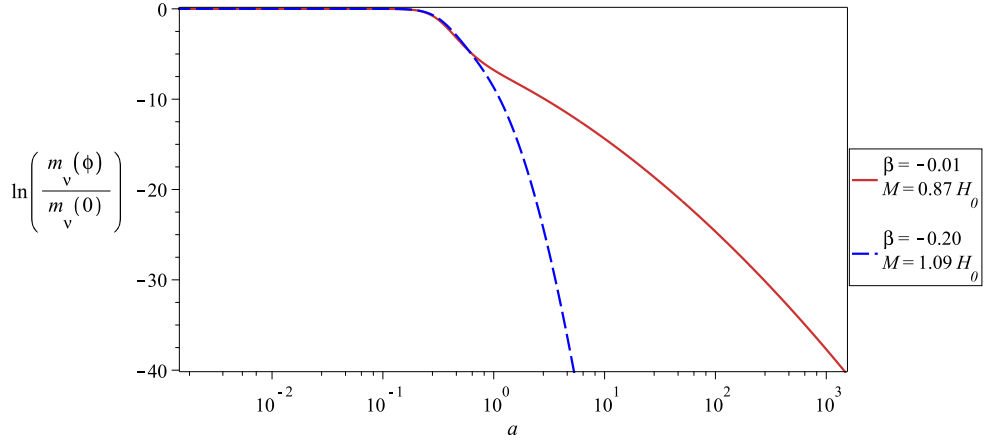}
	\caption{$\ln\left(\dfrac{m_\nu(\phi)}{m_\nu(0)}\right)$ in terms of $a$, for the parameters (\ref{parameters}) and initial conditions (\ref{ICs}).}
	\label{fig_m_nu}
\end{figure}

In figure \ref{fig_phi}, the evolution of the scalar field $\phi$ with respect to $\tau$ is plotted which starts from $\phi=10^{-8}$ and grows continually.
In figure \ref{fig_m_nu}, $\ln\left(\dfrac{m_\nu(\phi)}{m_\nu(0)}\right)$ is plotted with respect to the scale factor $a$ for $\{\beta=-0.01 (-0.20), M=0.87 H_0 (1.09 H_0)\}$. In the present era, i.e. $\tau\simeq0.93$($0.92$) (corresponding to $a=1$), $\phi(t)=0.72$ ($0.82$) and according to $m_\nu^*\simeq 0.0105 \text{ eV}$, for the current mass of neutrinos we have $m_\nu\simeq1.2\times10^{-5} \text{ eV}$ ($1.7\times10^{-6} \text{ eV}$).

\begin{figure}[H]
	\centering
	\includegraphics[height=7cm]{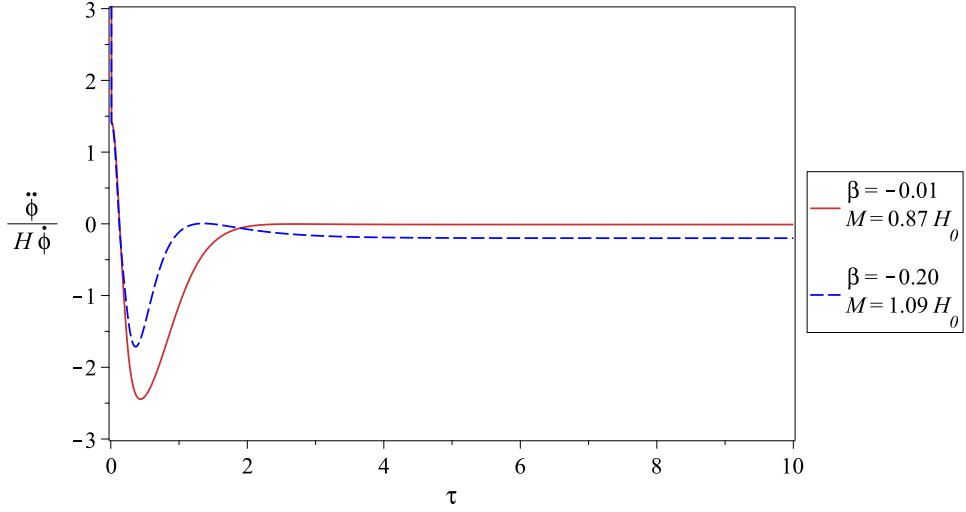}
	\caption{The rate of roll, i.e. $\dfrac{\ddot{\phi}}{H\dot{\phi}}$, in terms of $\tau$, for the parameters (\ref{parameters}) and initial conditions (\ref{ICs}).}
	\label{fig_beta}
\end{figure}

In figure \ref{fig_beta}, the rate of roll, i.e. $\dfrac{\ddot{\phi}}{H\dot{\phi}}$, is depicted in terms of $\tau$. As it can be seen, $\ddot{\phi}$ is positive initially which causes $\dot{\phi}$ to increase. Furthermore, as $\tau$ increases and the energy densities $\rho_\nu$, $\rho_m$ and $\rho_r$ decrease, the rate of roll tends to the $\beta$ parameter, as expected.

\begin{figure}[H]
	\centering
    \begin{subfigure}{0.45\textwidth}
		\centering
		\includegraphics[height=4.5cm]{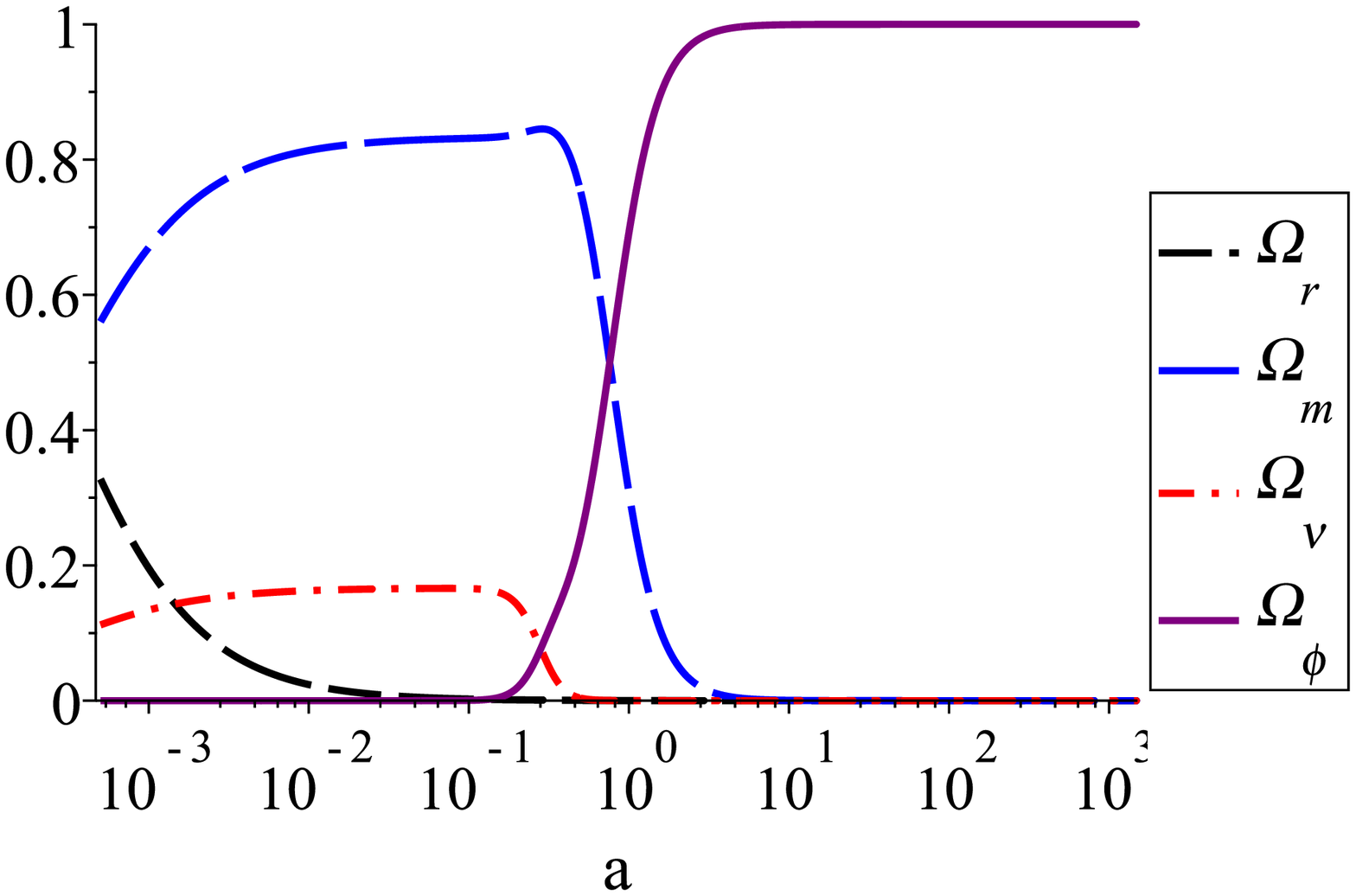}
		\caption{$\beta=-0.01, M=0.87 H_0$}
		\label{fig_Omega_0.01}
	\end{subfigure}
	\hfill
	\centering
    \begin{subfigure}{0.45\textwidth}
		\centering
		\includegraphics[height=4.5cm]{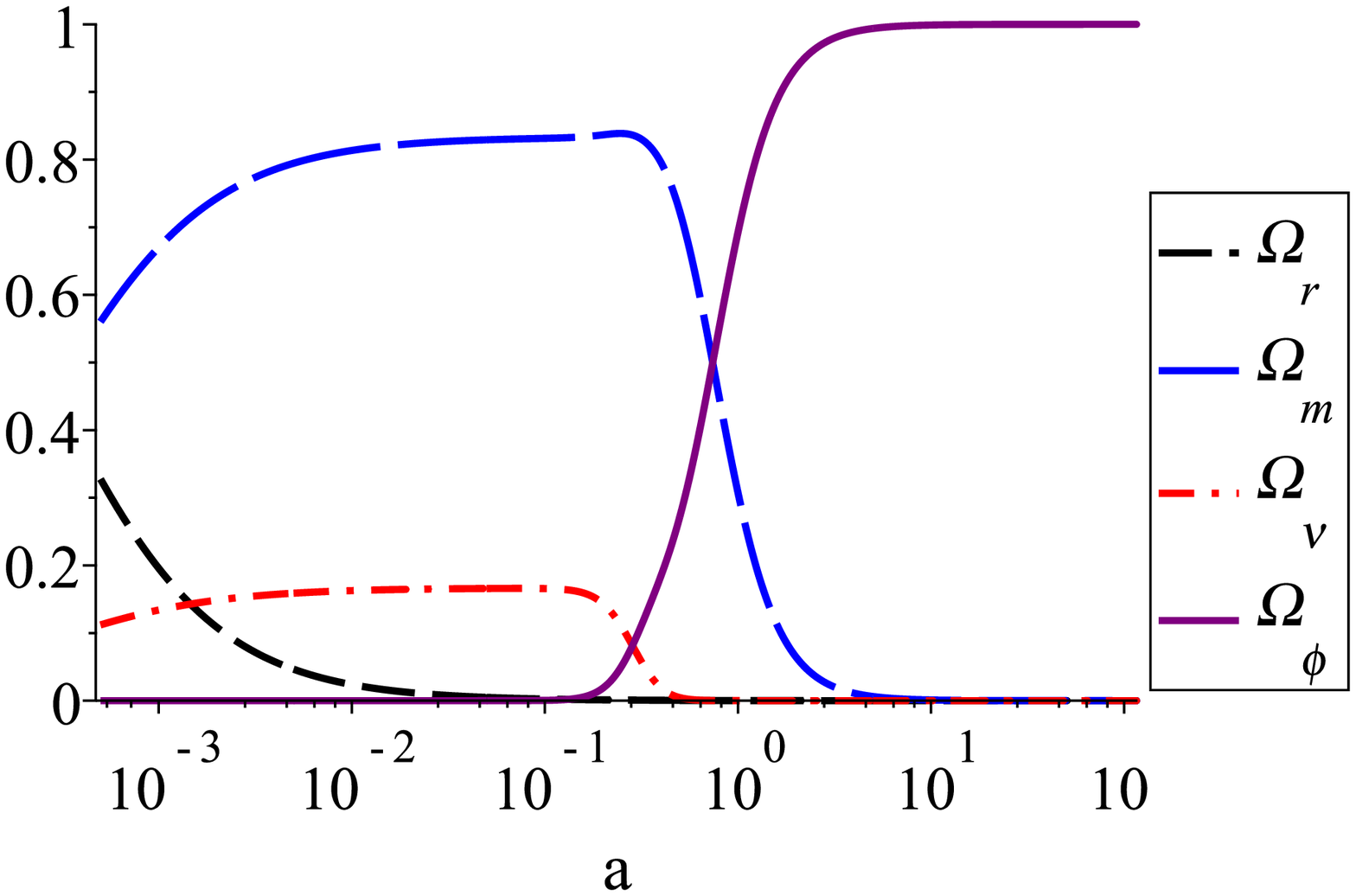}
		\caption{$\beta=-0.20, M=1.09 H_0$}
		\label{fig_Omega_0.2}
	\end{subfigure}
	\caption{Relative densities in terms of the scale factor $a$, for the parameters (\ref{parameters}) and initial conditions (\ref{ICs}).}
	\label{fig_Omega}
\end{figure}

In figure \ref{fig_Omega}, $\Omega_r$, $\Omega_m$, $\Omega_\nu$ and $\Omega_\phi$ which are respectively the relative densities of the radiation, the pressure-less matter, the neutrinos and the dark energy, are plotted in terms of the scale factor $a$ in the interval $\tau\in [0,10]$.

One can see in figure \ref{fig_Omega_0.01}(\ref{fig_Omega_0.2}), the Universe is initially at the matter-dominated era and then at $z\simeq0.32$($0.35$), it transits to the dark-energy-dominated era. According to the chosen parameters (\ref{parameters}) and initial conditions (\ref{ICs}), relative densities in the present era, i.e. $\tau\simeq0.93$($0.92$) (corresponding to $a=1$), are obtained as $\Omega_r = 0.00009$($0.00009$) , $\Omega_m = 0.308$($0.308$), $\Omega_\nu = 0.00007$($0.00001$) and $\Omega_\phi = 0.691$($0.691$) which are in the region expected by Planck 2015 data \cite{Planck}. Also, as we can see in figure \ref{fig_Omega} and as it was mentioned earlier, at any time $\tau\gtrsim1$, the scale factor of a universe with $\beta=-0.01$ is bigger than the scale factor of a universe with $\beta=-0.2$.

\section{Summary}
Using a scalar field, we have proposed a dynamical model for dark energy which describes the origin of positive late-time cosmic acceleration. Our model has $Z_2$ symmetry and the scalar field is initially located at the minimum of its potential ($\phi=0$). Unlike the symmetron model whose minimum of potential is non-zero \cite{symmetron}, we assume $V_\text{min}=0$ in our models. So, both the kinetic and potential energies of the scalar field are zero initially. Hence, the initial energy density of dark energy is zero, which alleviates the coincidence problem and is consistent with the present astrophysical data.

We have assumed a coupling between the scalar field and massive neutrinos. When neutrinos become non-relativistic ($m_\nu \simeq T_\nu$), this coupling causes the effective potential to become convex at $\phi=0$, the evolution of the scalar field begins and the $Z_2$ symmetry is broken. In other words, the onset of the scalar field's evolution is related to the neutrinos' mass through this coupling. This evolution results in the beginning of the cosmic acceleration at $z \simeq 0.6$ and the transition of the Universe from the matter-dominated era to the dark-energy dominated era at $z \simeq 0.3$.

Inspired by the inflationary constant-roll model, we have chosen the potential such that the evolution of the scalar field continues under the constant-roll condition ($\ddot{\phi}(t)=\beta H\dot{\phi}(t)$). In contrast to \cite{MaVaNs} in which the slow-roll condition is considered and the cosmic acceleration is not persistent, this choice of potential results in a persistent late-time acceleration and the deceleration parameter ($q$) tends to $-1-\beta$ eventually.
\\

As the scalar field settles down first at the fixed point $\phi=0$ and then by the symmetry breaking eventually evolves very slowly in a constant roll manner, the background evolution, as can be seen in figure \ref{fig_Omega}, is very similar to the $\Lambda CDM$ at the late time. This model can alleviate the coincidence problem, i.e. why the dark energy and dark matter densities are of the same order today or why dark energy density was negligible in earlier epochs. Our model, contrary to mass varying neutrino models with an adiabatic evolution of the scalar field, does not give rise to instabilities due to neutrino perturbation growing \cite{Afshordi,MaVaNs}.

Although it seems that by properly selecting the model parameters, the evolution of the background is similar to the $\Lambda CDM$, they may be distinguished from the evolution of perturbations.
The assumption of dynamical dark energy affects the matter perturbation evolution with respect to $\Lambda CDM$. This has been used to test dark energy models in the context of $\sigma_8$ tension (for example, see \cite{Lambiase}). Also, the massive neutrinos influence the evolution of the matter perturbation and the corresponding power spectrum \cite{Eisenstein}. The matter density perturbation is suppressed by the free-streaming massive neutrinos (see for example \cite{Geng}). For larger neutrino masses, the matter power spectrum suppression increases \cite{Lorenz}. As an outlook, one can study the imprint of the massive neutrinos,  interacting with the quintessence through the conformal coupling, on the evolution of the matter perturbation. The main difference with the usual mass varying neutrinos proposed before in the literature is that the early neutrino masses are larger than that in the late time, so we expect to encounter more power spectrum depression compared to the previous models.

\section*{Acknowledgement}
V. Anari likes to thank the Iran National Science Foundation (INSF) for the partial financial supports.


\vspace{2cm}
\begin{thebibliography}{}
\bibitem{acc1} A.G. Riess, et al. (Supernova Search Team Collaboration), Astron. J.
116, 1009 (1998), [arXiv:astro-ph/9805201]
\bibitem{acc2}S. Perlmutter, et al., (Supernova Cosmology Project Collaboration),
Astrophys. J. 517, 565 (1999), [arXiv:astro-ph/9812133]
\bibitem{acc3}L. Amendola and S. Tsujikawa, Dark Energy: Theory and Observations(Cambridge University Press, 2010)
\bibitem{wein1} S. Weinberg, Rev. Mod. Phys. 61, 1 (1989)
\bibitem{wein2}I. Zlatev, L.-M. Wang, P. J. Steinhardt, Phys. Rev. Lett. 82, 896 (1999)
\bibitem{wein3}V. Sahni and A. A. Starobinsky, Int. J. Mod. Phys.
D 9, 373 (2000), [arXiv:astro-ph/9904398]
\bibitem{ten1}T. Karwal, M. Kamionkowski, Phys. Rev. D 94, 10523 (2016)
\bibitem{ten2}M. C. González, Q. Liang, J. Sakstein, M. Trodden,  JCAP 04 (2021) 063, [arXiv: 2011.09895 [astro-ph.CO]]
\bibitem{ten3}M. Maziashvili, [arXiv:2111.07288 [astro-ph.CO]]
\bibitem{ten4}S. Vanozzi, Phys. Rev. D 102, 023518 (2020), [1907.07569 [astro-ph.CO]]
\bibitem{ten5}S. Vagnozzi, 	Phys. Rev. D 104, 063524 (2021), [2105.10425 [astro-ph.CO]]
\bibitem{quint1}R.R. Caldwell, R. Dave, P.J. Steinhardt, Phys. Rev. Lett. 80, 1582 (1998), [arXiv:astro-ph/9708069]
\bibitem{quint2}E. Elizalde, S. Nojiri, S.D. Odintsov, Phys. Rev. D 70, 043539 (2004), [arXiv:hep-th/0405034]
\bibitem{quint3} H.M. Sadjadi, M. Alimohammadi, Phys. Rev. D 74, 043506 (2006), [arXiv:gr-qc/0605143]
\bibitem{quint4}  H. M. Sadjadi,  Eur. Phys. J. C 66, 445 (2010), [arXiv:0904.13494[gr-qc]]
\bibitem{screen1} K. Hinterbichler, J.  Khoury,	Phys. Rev. Lett. 104, 231301 (2010)
\bibitem{screen2} H. M. Sadjadi, Phys. Dark. Univ. 22, 101 (2018), [arxiv[18033.05310[gr-qc]]
\bibitem{planckin1}Planck Collaboration, Y. Akrami et al., Planck 2018 results. X. Constaints on inflation, Astron. Astrophys. 641, A10 (2020), [arXiv:1807.06211 [astro-ph.CO]].
\bibitem{planckin2} C. Q. Geng, C. C. Lee, M. Sami, E. N. Saridakis, A. A. Starobinsky, JCAP 1706, 011 (2017), [arXiv[1705.01329]]
\bibitem{cr-2}J. Martin, H. Motohashi, T. Suyama, Phys. Rev. D 87, 023514 (2013), [arXiv:1211.0083]
\bibitem{constant-roll}H. Motohashi, A. A. Starobinsky, and J. Yokoyama, JCAP 1509, 018 (2015)
\bibitem{cr-1}H. Motohashi, A. A. Starobinsky, Europhys. Lett. 117, 39001 (2017), [arXiv:1702.05847]
\bibitem{cr0}H. Motohashi, A. A. Starobinsky, JCAP 11 (2019) 025, [arXiv:1909.10883]
\bibitem{cr1}M. Guerrero, D. R. Garcia, D. S. C. Gomez,  Phys. Rev. D 102, 123528 (2020)[ arXiv:2008.07260 [gr-qc]]
\bibitem{cr2}M. Shokri, J. Sadeghi, S. N. Gashti, Phys. Dark Univ. 35, 100923 (2022) [arXiv:2107.04756 [astro-ph.CO]]
\bibitem{cr3}V. K. Oikonomou, [arXiv:2106.10778[gr-qc]]
\bibitem{cr4}H. M. Sadjadi, N. Anari, Phys. Dark. Univ 27, 100474 (2020) [arXiv:1908.04266[gr-qc]]
\bibitem{shib1}T. Chiba, Phys. Rev .D 79, 083517 (2009), [arXiv:0902.4037 [astro-ph.CO]]
\bibitem{shib2}S. Dutta, R. J. Scherrer, Phys. Lett. B 704, 265 (2011) [arXiv:1106.0012 [astro-ph.CO]]
\bibitem{shib3} M. Honardoost, H. Mohseni Sadjadi, H. R. Sepangi, Gen. Rel. Grav. 48, 125 (2016), [arXiv:1508.06022 [gr-qc]]
\bibitem{action}R. Bean, E. E. Flanagan, and M. Trodden, Phys. Rev. D 78, 023009 (2008).
\bibitem{fardon}R. Fardon, A.E. Nelson and N. Weiner, JCAP 10 (2004) 005 [arXiv:0309800[astro-ph]].
\bibitem{MaVaNs}H. Mohseni Sadjadi and V. Anari, Phys. Rev. D 95, 123521 (2017) [arXiv:1702.04244[gr-qc]]
\bibitem{Brookfield}A. W. Brookfield, C. van de Bruck, D. F. Mota, and D. T. Valentini, Phys. Rev. D 73, 083515 (2006)[arXiv:0512367[astro-ph]]
\bibitem{Peccei}R. D. Peccei, Phys. Rev. D 71, 023527 (2005) [arXiv:0411137[hep-ph]].
\bibitem{Pettorino}V. Pettorino, N. Wintergerst, L. Amendola, and C. Wetterich, Phys. Rev. D 82, 123001 (2010)[arXiv:1009.2461 [astro-ph.CO]].
\bibitem{Mota}A.W. Brookfield, C. van de Bruck, D.F. Mota and D. Tocchini-Valentini, Phys. Rev. D 73, 083515 (2006) 083515, [arXiv:0512367 [astro-ph]]
\bibitem{Pietroni}M. Pietroni, Phys. Rev. D 72, 04235 (2005), [arXiv:0505615 [astro-ph]].
\bibitem{JCAP}H. M. Sadjadi, V. Anari,  JCAP, 10 (2018)036, [arXiv:1808.01903[gr-qc]]
 \bibitem{JCAP1} M. Sami, S. Myrzakul, M. Al Ajm, 	[arXiv:1912.12026 [gr-qc]]
\bibitem{Planck}Planck collaboration, P.A.R. Ade et al., Planck 2015 results. XIII. Cosmological parameters, Astron. Astrophys. 594, A13(2016)  [arXiv:1502.01589 [astro-ph.CO]] .
\bibitem{symmetron}K. Hinterbichler, J. Khoury, A. Levy, A.  Matas, Phys. Rev.D, 84, 103251 (2011), [arXiv:1107.2112[astro-ph.CO]].
\bibitem{Pastor}J. Lesgourgues, S. Pastor, Adv. High Energy Phys. 2012 (2012) 608515, [arXiv:1212.6154 [hep-ph]].
\bibitem{Planck 2018}Planck Collaboration, N. Aghanim, et al., Planck 2018 results. VI. Cosmological parameters, arXiv:1807.06209.
\bibitem{Gariazzo}S. Gariazzo, P. F. de Salas, S. Pastor, JCAP 07 (2019) 014, [arXiv:1905.11290 [astro-ph.CO]].
\bibitem{Feng}Lu Feng, Rui-Yun Guo, Jing-Fei Zhang, Xin Zhang, Phys. Lett. B 827 (2022) 136940.
\bibitem{Afshordi}[N. Afshordi, M. Zaldarriaga, and K. Kohri, Phys. Rev. D 72, 065024 (2005).
\bibitem{Lambiase}G. Lambiase, S. Mohanty, A. Narang,  P. Parashari, Eur. Phys. J. C (2019) 79:141.
\bibitem{Eisenstein}W. Hu, D. J. Eisenstein, and M. Tegmark, Phys.Rev.Lett.80:5255-5258,1998, [[arXiv:astro-ph/9712057].
\bibitem{Geng}Chao-Qiang Geng, Chung-Chi Lee, R. Myrzakulov, M. Sami, Emmanuel N. Saridakis,   JCAP 01 (2016) 049, [arXiv:1504.08141 [astro-ph.CO]].
\bibitem{Lorenz}C.  S. Lorenz, L. Funcke, M.  Löffler, E. Calabrese,    Phys. Rev. D 104, 123518  (2021), [arXiv:2102.13618 [astro-ph.CO]].
\end{thebibliography}
\end{document}